\documentclass[12pt,letterpaper]{JHEP3}
\usepackage{cite}
\newcommand{\beq}{\begin{equation}}
\newcommand{\eeq}{\end{equation}}
\newcommand{\beqa}{\begin{eqnarray}}
\newcommand{\eeqa}{\end{eqnarray}}
\newcommand{\beqar}{\begin{eqnarray*}}
\newcommand{\eeqar}{\end{eqnarray*}}

\newcommand{\reef}[1]{(\ref{#1})}

\newcommand{\norm}[1]{\raise.3ex\hbox{:}#1\raise.3ex\hbox{:}}

\newcommand{\gsim}{\mathrel{\raisebox{-.6ex}{$\stackrel{\textstyle>}{\sim}$}}}
\newcommand{\lsim}{\mathrel{\raisebox{-.6ex}{$\stackrel{\textstyle<}{\sim}$}}}

\newcommand{\Om}{\Omega}

\renewcommand{\L}{\Lambda}

\newcommand{\cF}{{\cal F}}

\newcommand\tA{{\widetilde A}}
\newcommand\tB{{\widetilde B}}

\newcommand\tL{{\widetilde L}}

\newcommand\N{{N}}

\def\tf#1#2{{\textstyle{#1 \over #2}}}

\def\overleftrightarrow#1{\vbox{\ialign{##\crcr
     $\leftrightarrow$\crcr\noalign{\kern-0pt\nointerlineskip}
     $\hfil\displaystyle{#1}\hfil$\crcr}}}

\def\lsim{\mathrel{\mathstrut
\smash{\ooalign{\raise2.5pt\hbox{$<$}\cr\lower2.5pt\hbox{$\sim$}}}}}
\def\gsim{\mathrel{\mathstrut
\smash{\ooalign{\raise2.5pt\hbox{$>$}\cr\lower2.5pt\hbox{$\sim$}}}}}

\def\sqr#1#2{{\vcenter{\vbox{\hrule height.#2pt
         \hbox{\vrule width.#2pt height#1pt \kern#1pt
            \vrule width.#2pt}
         \hrule height.#2pt}}}}
\def\square{\mathop{\mathchoice\sqr56\sqr56\sqr{3.75}4\sqr34\,}\nolimits}

\title{Unbounded entropy in spacetimes with positive cosmological
constant}

\author{
Raphael Bousso and Oliver DeWolfe \\
Institute for Theoretical Physics\\
University of California, Santa Barbara, CA 93106-4030, U.S.A.\\
E-mail: \email{bousso@itp.ucsb.edu}, \email{odewolfe@itp.ucsb.edu}}

\author{
Robert C. Myers\thanks{On leave from McGill University.}\\
Perimeter Institute for Theoretical Physics\\
35 King Street North, Waterloo, Ontario N2J 2W9, Canada;\\
Department of Physics, University of Waterloo\\
Waterloo, Ontario N2L 3G1, Canada;\\
Department of Physics, McGill University\\
Montr\' eal, Qu\' ebec H3A 2T8, Canada\\
E-mail: \email{rmyers@perimeterinstitute.ca}}

\abstract{ In theories of gravity with a positive cosmological
constant, we consider product solutions with flux, of the form
(A)dS$_p\times S^q$.  Most solutions are shown to be perturbatively
unstable, including all uncharged dS$_p\times S^q$ spacetimes.  For
dimensions greater than four, the stable class includes universes
whose entropy exceeds that of de~Sitter space, in violation of the
conjectured ``$N$-bound''.  Hence, if quantum gravity theories with
finite-dimensional Hilbert space exist, the specification of a
positive cosmological constant will not suffice to characterize the
class of spacetimes they describe.  }

\preprint{\hepth{0205080} \\ NSF-ITP-02-018}
\begin{document}

\section{Introduction} \label{intro}

String theory has had some success in defining quantum gravity in
certain asymptotically flat and asymptotically Anti-de~Sitter (AdS)
spacetimes.  It is natural, therefore, to seek a fundamental
description of asymptotically de~Sitter spacetimes.  This is
particularly interesting because our own universe may belong to this
class~\cite{Rie98,Per98}.  Moreover, de~Sitter space is a simple arena
in which one may confront conceptual problems that arise in the
description of any cosmological spacetime.

Although some progress has been reported (see, e.g.,
Refs.~\cite{Sil01,Hul01,GutStr02}), there is presently no fully
satisfactory embedding of de~Sitter space into string
theory.\footnote{Some recent approaches to quantum gravity in
de~Sitter space do not use string theory as a starting point (e.g.,
Refs.~\cite{Str01,Wit01}; see also further references in
Refs.~\cite{BalDeb01,SprVol01}).}  It is important to understand
whether this is only a technical problem, or whether significant new
developments (comparable, e.g., to the discovery of
D-branes~\cite{Pol95}) will be required for progress.  To this end,
one may deduce properties of de~Sitter quantum gravity by applying the
general consequences we expect from a complete quantum theory, namely
semi-classical gravity and the holographic
principle~\cite{Tho93,Sus95,Bou99b,Bou99c}.  One can then ask
whether string theory may display such properties, and specific
challenges may be pinpointed or obstructions identified.

One recent line of thought, initiated by Banks and
Fischler~\cite{Fis00a,Fis00b,Ban00}, centers on the semi-classical
result that de~Sitter space has a finite entropy
$S_0$~\cite{GibHaw77a}, inversely related to the cosmological
constant.  This approach reasons that de~Sitter space should be
described by a theory with a finite number $e^{S_0}$ of independent
quantum states.  Any additional states would be superfluous, as they
would render the theory more complex than the phenomena it
describes---an uneconomical and arbitrary excess.\footnote{Such an
excess should be distinguished from the situation of an ordinary gauge
symmetry, where the variables are redundant, but a quotient
nonetheless produces a minimal physical Hilbert space.}  Hence, it is
asserted, one must construct a quantum gravity theory with a finite
dimensional Hilbert space in order to describe de~Sitter space.

In its strongest form, this reasoning leads to a new perspective on
the origin of vacuum energy~\cite{Ban00}: the ``$\Lambda$-$N$
correspondence''~\cite{Bou00a}.  The cosmological constant $\Lambda$
should be understood as a direct consequence of the finite number of
states $e^N$ in the Hilbert space describing the world.  $\Lambda$
effectively provides a cutoff on observable entropy, ensuring that the
theory need never describe phenomena requiring a larger number of
states; the smaller the Hilbert space, the larger the cosmological
constant.

Originally these observations were made in the context of spacetimes
that asymptote to de~Sitter space in the
future~\cite{Fis00a,Fis00b,Ban00}.  However, for reasons first noted
in Ref.~\cite{Bou00a} and further elaborated below, it would be
unnatural for any particular theory to describe {\em only\/}
spacetimes of this type.  A positive cosmological constant does not
guarantee the presence of de~Sitter asymptotic regions; worse, their
existence can be affected by small deformations of Cauchy data.
Consequently, a larger class of spacetimes must be identified as the
``gravity dual'' if theories with finite Hilbert spaces are to play a
role in quantum gravity.

Once asymptotic conditions are abandoned, the finiteness of entropy is
no longer obvious.  A necessary condition for a (coarse-grained)
spacetime to be described by a theory with $e^N$ states is that the
entropy accessible to any observer in that spacetime must obey
\begin{equation}
S\leq N.
\label{eq-sleqn}
\end{equation}
It may be possible to characterize various different suitable
classes of spacetimes containing the asymptotically de~Sitter
spacetimes as a subset.  However, in view of the ``$\Lambda$-$N$
correspondence'', the simplest proposal is to consider the set
$\mathbf{all}(\Lambda(N))$: the class of spacetimes with positive
cosmological constant $\Lambda(N)$, irrespective of asymptotic
conditions and types of matter present~\cite{Bou00a}.

Generally, one expects a good quantum gravity theory to predict
specific matter content.  However, of course, we do not yet have a quantum
gravity theory with a finite number of states, and so we are unable to
determine which matter fields should be allowed.  This is a drawback,
because restrictions on the matter content might enforce the condition
(\ref{eq-sleqn}) when it would otherwise be violated.  However, the
covariant entropy bound~\cite{Bou99b} limits entropy in terms of a
geometric feature (area), assuming nothing about the matter content
except that fairly generic energy conditions hold.  Thus, if the
geometries in a certain class of spacetimes obey suitable
restrictions, Eq.~(\ref{eq-sleqn}) may follow from the holographic
principle with few or no detailed assumptions about matter (see
Ref.~\cite{Bou02} for a review).

By applying the condition (\ref{eq-sleqn}) to the candidate set
$\mathbf{all}(\Lambda(N))$, one obtains a consistency requirement, the
``$N$-bound'', which we state here for $D=4$~\cite{Bou00a}: {\em In
any universe with a positive cosmological constant $\L$ (as well as
arbitrary additional matter that may well dominate at all times) the
observable entropy $S$ is bounded by $\N{=}3\pi/\L$.}  (We use Planck
units throughout.)  For the special case of central observers in
spherically symmetric spacetimes with $\Lambda(N)$, the $N$-bound was
proven~\cite{Bou00a} using the covariant entropy
bound~\cite{Bou99b,Bou02}.  This proof relies on general geometric
properties of spacetimes with positive cosmological constant, and on a
bound on matter entropy in de~Sitter space~\cite{Bou00b}.  It applies
for all $D\geq 4$.

In this paper, however, we show that counterexamples to the $N$-bound
exist in $D>4$.  They are not counterexamples to the covariant entropy
bound, but they evade the proof of Ref.~\cite{Bou00a} by violating the
assumption of spherical symmetry.  Their key novel ingredient is flux,
which is used to stabilize a product metric with one large or
non-compact factor.

This shows, in particular, that the mere specification of a positive
cosmological constant does not suffice to guarantee finite observable
entropy.  $\Lambda$ cannot be in correspondence with $N$ unless some
additional conditions hold that exclude our counterexamples.  In
particular, conditions on matter content may be required.  This leaves
open the question of whether finite Hilbert space theories can have a
gravity dual.


In Sec.~\ref{sec-ds}, we discuss the classification and entropy of
solutions related to de~Sitter space.  We formulate conditions that
the gravity dual of a finite Hilbert space theory should satisfy (if
it exists).

In Sec.~\ref{SolutionSec}, we consider a $D=p+q$ dimensional theory
with positive cosmological constant and a $q$-form field strength.  In
addition to the dS$_{p+q}$ solution, we present all solutions of the
product form (A)dS$_p\times S^q$ with or without $q$-form flux on the
$S^q$.  We note that these solutions can be generalized to include a
black hole; this will later permit us to introduce entropy into the
AdS$_p\times S^q$ solutions.

In Sec.~\ref{StableSec}, we investigate the stability of the product
solutions we have described.  Stability is essential if we wish to
interpret their horizon areas as entropy.  Our results apply for $p >
2$. We find that stable dS$_p \times S^q$ solutions exist for $2 \leq
q \leq 4$ and certain values of flux.  Furthermore, stable AdS$_p
\times S^q$ solutions exist for $q=2,3$ with any flux, and for other
even $q$ with flux sufficiently large.  We further argue that if an
AdS$_p\times S^q$ solution is stable, all Schwarzschild-AdS$_p\times
S^q$ solutions with the same asymptotic conditions will also be
stable.

Finally, in Sec.~\ref{EntropySec} we show that some of the stable
solutions we have described violate the $N$-bound.  For $p>2$, the
stable AdS$_p\times S^q$ solutions permit Schwarzschild black holes of
arbitrarily large entropy.  Also for $p>2$, the cosmological horizon
area of the dS$_p\times S^q$ spacetimes can exceed that of dS$_{p+q}$.
For $D>4$, this demonstrates that the observable entropy can be
arbitrarily large at fixed positive $\Lambda$.

We present our conclusions in Sec.~\ref{discuss}.

\section{Classification and properties of $\Lambda>0$ spacetimes}
\label{sec-ds}

\subsection{Asymptotically de~Sitter spacetimes}

To say that a spacetime is asymptotically de~Sitter is an ambiguous
statement.  Unlike the flat and AdS cases, de~Sitter space has two
disconnected infinities, one in the past and one in the future.  It is
useful to distinguish which of these infinities are present.  (For a
brief summary of the de~Sitter geometry, see the appendix
of Ref.~\cite{Bou00a}.)

For a given cosmological constant $\Lambda>0$, let us call a spacetime
dS$^\pm$ if it possesses both infinities.  An example is a small
perturbation about de~Sitter space, or the Schwarzschild-de~Sitter
solution.  The $\mathbf{dS^\pm}(\Lambda)$ set of spacetimes is the
intersection of the $\mathbf{dS^+}$ and $\mathbf{dS^-}$ sets, i.e.,
the spacetimes that asymptote to de~Sitter space\footnote{It is useful
to characterize such spacetimes without making reference to all of
${\cal I}^+$.  As a working definition of $\mathbf{dS^+}$, we propose
the following.  Recall that the causal diamond $C(p,q)$ is the
intersection of the future of $p$ with the past of $q$.  A spacetime
is in $\mathbf{dS^+}$ if it contains at least one worldline $\gamma$
with the following two properties.  1.~Let $\tau$ be the proper time
on $\gamma$.  $\gamma$ is geodesic after some finite time $\tau_0$.
2. Let $p$, $q$ be points on the world line such that $\tau(p) <
\tau(q)$.  The geometry of the causal diamonds $C(p,q)$ asymptotes to
the static patch of de~Sitter space as $\tau(p) \to
\infty$, $\tau(q)-\tau(p) \to \infty$.  An analogous definition
applies for $\mathbf{dS^-}$.} in the future or in the past, respectively. For
example, if our universe started from a big bang and is evolving to a
de~Sitter state, it belongs to the $\mathbf{dS^+}$ category.  Finally,
we may define an even larger set, $\mathbf{all}(\Lambda)$, by
specifying only the cosmological constant $\Lambda$ (defined here to be the
lowest attainable vacuum energy in the theory), without
demanding any asymptotic conditions at all.  For example, a closed
recollapsing universe can have a positive vacuum energy as long as the
matter energy always dominates.

Furthermore, small changes in the Cauchy data can take a spacetime in
and out of any of the $\mathbf{dS^+}$, $\mathbf{dS^-}$, and
$\mathbf{dS^\pm}$ classes.  This is in further contrast to
asymptotically flat or AdS universes, which retain their asymptotic
structure independently of the matter injected at the boundary (unless
the cosmic censorship conjecture is violated), and independently of
continuous deformations of Cauchy data.

For example, suppose we lived in a dS$^\pm$ universe.  Outside our
past light cone, a shell of matter might be collapsing toward us.
When it reaches our causal domain, it will form a
Schwarzschild-de~Sitter black hole, which will evaporate and leave
de~Sitter space behind.  However, black holes larger than the
cosmological horizon cannot exist in de~Sitter space.  Consider a
slightly subcritical shell, which forms a nearly-maximal black hole.
If a tiny amount of matter is added to the shell initially, the
collapse will result in a big crunch.  A large enough black hole
cannot form; instead, there will be contracting time slices on which
the matter density dominates over the vacuum energy.  Hence the
universe collapses entirely, which places it in the $\mathbf{dS^-}$
class.

A second example is that of a $\Lambda >0$ closed universe filled with
dust.  Starting from a big bang, such a universe may expand
indefinitely (dS$^+$), or it may recollapse, depending on the ratio of
dust to vacuum energy.  Thus, two universes with different fates can
possess early time Cauchy surfaces with almost identical matter
density and expansion rate.

\subsection{$\mathbf{dS^+}$ and finite entropy}

In a dS$^+$ spacetime, any observer is surrounded by a cosmological
event horizon, whose area is inversely related to the cosmological
constant, $\Lambda$.  Ordinary entropy is lost to the observer when
matter crosses this horizon.  Just as with a black hole horizon, in
order to maintain a generalized second law of
thermodynamics~\cite{Bek72,Bek73,Bek74} the horizon must be assigned
an entropy equal to a quarter of its area, $A_{\rm
hor}$~\cite{GibHaw77a}:
\begin{equation}
S_{\rm hor} = \frac{A_{\rm hor}}{4} \,.
\label{eq-bekhaw}
\end{equation}
The horizon area generally varies with the amount of matter enclosed.
With reasonable assumptions about the maximal entropy of
matter~\cite{Bek81,SchBek89}, one finds that the horizon area adjusts
itself such that the combined entropy of matter and horizon is less
than that of empty de~Sitter space~\cite{Bou00b}:
\begin{equation}
S_{\rm matter}+ S_{\rm hor} \leq S_0 \,.
\label{eq-ssn}
\end{equation}
This result also follows immediately from the generalized second law.
The final state of the system is empty de~Sitter space, and earlier
configurations may not exceed its entropy.

The entropy $S_0$ of empty de~Sitter space is determined by the
cosmological constant alone:
\begin{equation}
S_0 = {\Omega_{D-2}\over4}
\left[{(D-1)(D-2)\over2\L}\right]^{D-2\over2}\
, \label{general}
\end{equation}
where $\Omega_{D-2}=2\pi^{(D-1)/2}/\Gamma((D-1)/2)$ is the area of a
unit ($D$--2)-sphere.  This is the largest entropy (counting both
horizons and matter) observable in any of the dS$^+$
spacetimes.\footnote{A future event horizon is present in all
expanding homogeneous isotropic universes with equation of state
$p=w\rho$, $-1 \leq w < 1/3$ (for $D=4$) (see, e.g.,
Refs.~\cite{HelKal01,FisKas01}).  However, this horizon has finite
maximal area only in the de~Sitter case ($w=-1$).  For the remaining
values of $w$ the horizon area grows without bound.  Hence there is no
bound on observable entropy in ``quintessence'' models, and they
cannot be dual to a theory with a finite number of states.}

Note, however, that $S_{\rm matter}$ includes only entropy that is
causally accessible to the observer, i.e., entropy contained in causal
diamond regions.  (Generally, a causal diamond is the intersection of
the past and the future of an observer's world line~\cite{Bou00a}.)
This is sensible because unobservable regions are operationally
meaningless.  Indeed, the unitary evolution of black holes suggests
that causally disconnected observers cannot be simultaneously
described without encountering a paradox~\cite{SusTho93}.  The
restriction to causal diamonds is crucial for the validity of
Eq.~(\ref{eq-ssn}), as the global entropy may be formally
unbounded.\footnote{With the stronger assumption that the spacetime is
dS$^\pm$, Eq.~(\ref{general}) is a bound on the matter entropy on
initial global slices~\cite{Bou02}.}

Analogously to the case of black holes, the entropy of dS$^+$
universes may be interpreted microscopically in terms of the number of
accessible quantum states.  (Though well motivated by the D-brane
model of black hole entropy~\cite{StrVaf96}, we must stress that this
interpretation is not inevitable.)  By Eq.~(\ref{eq-ssn}), it follows
that $e^{S_0}$ states suffice to describe all dS$^+$ universes with
cosmological constant $\Lambda$.

\subsection{Defining gravity sectors with finite entropy}
\label{DefiningSec}

The de~Sitter horizon differs from the horizon of a black hole in an
important respect.  The mass of a Schwarzschild black hole determines
the area of its horizon and thus the size of the Hilbert space
describing the black hole; similarly, the cosmological constant sets
the area of the de~Sitter horizon.  However, the black hole mass is a
variable parameter of the solution, which can even be time dependent;
whereas the cosmological constant $\Lambda$ (defined as the lowest
accessible vacuum energy) can be interpreted as a fundamental
characteristic of the theory.

In the introduction we argued that a theory should not contain more
states than the phenomena it describes.  This suggests that the dS$^+$
class of universes is in fact fully described by a quantum gravity
theory with a Hilbert space ${\cal H}$ of finite
dimension~\cite{Fis00a,Fis00b,Ban00}
\begin{equation}
\dim {\cal H} = e^{S_0}.
\end{equation}
For any quantum system, let us define $N$, the {\em number of degrees
of freedom}, to be the logarithm of the dimension of the Hilbert
space.  (With this definition, degrees of freedom are spin-like,
rather than fields or harmonic oscillators.)  Then we may restate our
conclusion as follows.  A quantum description of the dS$^+$ universes
requires only a finite number of degrees of freedom,
\begin{equation}
N= S_0.
\label{Nbound}
\end{equation}
Of course, it is quite possible that a theory of dS$^+$ universes will
make use of a larger, or infinite-dimensional Hilbert space.  It is
remarkable, however, that a finite Hilbert space is in principle
sufficient.  By contrast, such a Hilbert space would necessarily be
insufficient to describe asymptotically flat or AdS spacetimes, which
can contain arbitrarily large entropy (for example, in the form of
large black holes).

This result may point at a new class of theories, distinct from those
for asymptotically flat or AdS spacetimes.  The finiteness of $N$ may
be a crucial qualitative feature underlying a successful description
of dS$^+$ spacetimes.

Suppose, however, that we were given a theory ${\cal T}$ with a finite
number of degrees of freedom, $N$, and that this theory described all
dS$^+$ universes with the corresponding value of $\Lambda$,
\begin{equation}
\Lambda(N) = {(D-1)(D-2)\over 2} \left( {\Omega_{D-2}\over 4N}
\right)^{2\over D-2}.
\label{eq-ln}
\end{equation}
We discussed earlier that the difference in initial conditions between
a spacetime ${\cal M}_1$ in the dS$^+$ class, and a spacetime ${\cal
M}_2$ with no asymptotic de~Sitter regions, can be arbitrarily small.
It would be unnatural for a small deformation to invalidate the
description of the spacetime by the theory ${\cal T}$.  Yet, in the
absence of a future asymptotic region, the second law argument leading
to Eq.~(\ref{eq-ssn}) cannot be completed.\footnote{There are
additional reasons why reference to an asymptotic region will not be
ultimately satisfactory.  Classically, a dS$^+$ spacetime may contain
observers who do not get to ${\cal I}^+$---for example, if it contains
black holes.  Still worse, at the semi-classical level it becomes
clear that {\em no\/} observer gets to ${\cal I}^+$.  No system can
withstand the thermal radiation~\cite{GibHaw77a} permeating de~Sitter
space indefinitely; rather, it will be thermalized in a finite time.
Another aspect of this problem comes from noting that black holes will
also form spontaneously in the thermal radiation~\cite{BouHaw98}.  It
follows that eventually, after a long but finite time, any observer
will be swallowed up by a black hole, with a finite probability
(approaching one).  In the case of a black hole of nearly maximal
size, this process is indistinguishable from a big crunch.  The
creation rate for such black holes is roughly
$e^{-S_0/3}$~\cite{GinPer83}. Note that this is even greater than the
estimated rate of Poincar\'e recurrences, $e^{-S_0}$~\cite{DysLin02},
which independently limits observer lifetimes in de~Sitter space.
More stringent upper bounds would be desirable but are not needed
here.}  It is not clear if there is an upper bound to the entropy in
${\cal M}_2$.  Hence, it is unclear if ${\cal T}$ contains enough
degrees of freedom to fully describe ${\cal M}_2$.

This poses a challenge.  For a given value of $N$, one would like to
identify a set of spacetimes ${\cal C}(N)$ that could be described by
a theory ${\cal T}$ whose Hilbert space has dimension
$N$.\footnote{Neither ${\cal T}$ nor ${\cal C}(N)$ need to be unique
for given $N$.  We do not assume this, nor do we assume that a
consistent gravity theory exists for every value of $N$.  Moreover, we
must stress that the same low energy Lagrangian may contain different
sectors of solutions corresponding to different values of $N$,
including $N=\infty$.  Each of these sectors would be described by a
different theory ${\cal T}$.  (This is familiar from the AdS/CFT
correspondence.  Solutions differing only by the number $n$ of units
of flux are described by different $SU(n)$ gauge theories.)}  Such a
class should possess the following three properties:
\begin{enumerate}
\item{${\cal C}(N)$ should include, but not be limited to, the dS$^+$
solutions with $\Lambda = \Lambda(N)$:
\begin{equation}
{\cal C}(N) \supset
\mathbf{dS^+}(\Lambda(N)).
\end{equation}}
\item{${\cal C}(N)$ should be closed under smooth deformations of
Cauchy data.}
\item{${\cal C}(N)$ must not include any universes with observable
entropy greater than $N$.}
\end{enumerate}

Note that only the last condition is strictly necessary.  As discussed
in Sec.~\ref{intro}, one expects a unified theory to predict a
particular matter content.  This will limit the range of spacetimes
defined by the first two conditions, and thus it may prevent
violations of the third which a more ignorant analysis would produce.
It may even remove the need for the second condition altogether.

For now, we have no information about matter content and will assume
only that reasonable energy conditions are satisfied.  Then the above
conditions are useful for a preliminary assessment of the potential
role of finite Hilbert space theories in quantum gravity.

Neither the existence nor the uniqueness of ${\cal C}(N)$ is
guaranteed {\em a priori}.  One could exclude its existence, e.g., by
showing that dS$^+$ Cauchy data can be smoothly deformed to yield a
universe with observable entropy greater than $N$.  This would render
the case for finite Hilbert space theories related to quantum gravity
in de~Sitter space less compelling.  If a suitable set is found,
however, it will enhance the motivation for constructing such
theories.

A natural proposal~\cite{Bou00a} would be to choose ${\cal C}(N)
=\mathbf{all}(\Lambda(N))$, the set of all universes with positive
cosmological constant $\Lambda(N)$ given by Eq.~(\ref{eq-ln}).  Since
the cosmological constant is not part of the variable Cauchy data,
this choice manifestly satisfies the first two properties.  However,
we find that the third property is violated by some spacetimes in
$\mathbf{all}(\Lambda(N))$.  We now turn to a description of these
solutions.

\section{Product spacetimes with flux}

\label{SolutionSec}

We consider solutions to the following action in $D=p+q$ dimensions:
\begin{equation}
I={1\over 16\pi}\int
d^{p+q}x\sqrt{-g}\left(R-2\L-{1\over2\,q!}F_q^2\right). \label{actpq}
\end{equation}
(Recall that are working in Planck units with $G_D=1$.)
This describes Einstein gravity with a cosmological constant $\L$,
taken to be positive, coupled to a $q$-form field strength, $F_q=d
A_{q-1}$.  The equations of motion may be written as
\begin{eqnarray}
\label{EinsteinEqn}
R_{MN} &=& {1 \over {2 (q-1)!}} F_{M P_2 \cdots P_q} F^{\;\; P_2
\cdots P_q}_N - {{(q-1)} \over {2 (D-2) \, q!}} g_{MN}F_{q}^2 + {2
\over D-2} \, g_{MN} \Lambda \,, \\ d * F_q &=& 0 \,.
\label{MaxwellEqn}
\end{eqnarray}
The most symmetric solution is $d$-dimensional de~Sitter space, with
$F=0$ and metric
\begin{equation}
ds^2= - V(r) \, dt^2 + \frac{1}{V(r)} dr^2 + r^2 d\Omega_{D-2}^2,
\label{dSmetric}
\end{equation}
where
\begin{equation}
V(r) = 1 - {r^2\over L^2}.
\end{equation}
The radius of curvature is given in terms of the cosmological constant
by
\begin{equation}
L^2={(D-1)(D-2) \over 2 \Lambda}.
\end{equation}

Next, we find product solutions of the form $K_p \times M_q$, where
$K_p$ is Lorentzian with coordinates $x^\mu$, $M_q$ is Riemannian with
coordinates $y^\alpha$, and both factors are Einstein:
\begin{eqnarray}
\label{BackgroundCurv}
R_{\mu\nu} = {p-1 \over L^2} g_{\mu\nu} \,, \quad \quad R_{\alpha
\beta} = {q-1 \over R^2} g_{\alpha \beta} \,.
\end{eqnarray}
Additionally, we take the field strength $F_q$ to be proportional to
the volume form on $M_q$:
\begin{eqnarray}
\label{BackgroundFlux}
F_q = c\, {\rm vol}_{M_q}\,,
\end{eqnarray}
where ${\rm vol}_{S^q}$ is normalized so that $\oint {\rm vol}_{S^q} =R^q \,
\Om_q$.  This field strength automatically satisfies the Maxwell equation
(\ref{MaxwellEqn}) and the Bianchi identity $dF=0$.  Einstein's
equations now permit a family of solutions,
parametrized by the dimensionless flux
\begin{eqnarray}
\cF \equiv { c^2 \over 4 \Lambda} \,.
\end{eqnarray}
The curvature radii $L$, $R$ satisfy
\begin{eqnarray}
{p-1\over L^2}={ 2 \Lambda \over D-2}
\left[1- (q-1){\cal F}\right]\ , \quad \quad
{q-1\over R^2}= { 2 \Lambda \over D-2}
\left[1+ (p-1){\cal F}\right]\ .  \label{constrainpq}
\end{eqnarray}
Since $ \cF > 0$ by definition and we assume $\Lambda >0$, equation
(\ref{constrainpq}) requires $R^2 >0$ as well.  Hence, $M_q$ must have
positive curvature. We will generally take $M_q = S^q$, the
$q$-dimensional sphere.  As we discuss in the next section, this is
the choice most likely to be stable.

On the other hand, we see that $L^2$ has indefinite sign.  For small
$\cF$, one finds $L^2>0$.  This means that $K_p$ is positively curved
and can be taken to be dS$_p$.  At the value
\begin{eqnarray}
\label{Fm}
\cF = \cF_m \equiv {1 \over q-1} \,,
\end{eqnarray}
the curvature radius diverges.  In this case $K_p$ is flat; it can be
taken to be $p$-dimensional Minkowski space, for example.  For $\cF >
\cF_m$, $L^2$ becomes negative.  This corresponds to a change of sign
of the Ricci scalar.  The Lorentzian factor will be negatively curved,
and it is useful to define $\tL^2 \equiv - L^2$.  We can take $K_p$ to
be $p$-dimensional anti-de Sitter space (with real curvature radius
$\tL$) in this case.  Note that $\tL$
satisfies
\begin{eqnarray}
{p-1\over \tL^2}&=&{ 2 \Lambda \over D-2} \left[ (q-1){\cal F} -1
\right] \,.
\end{eqnarray}
We observe that as $\L\rightarrow0$ (with $c$ fixed), $R$ and $\tL$ remain
finite and our solutions reproduce the usual Freund-Rubin compactification
\cite{FreRub80} with geometry AdS$_p\times S^q$.

Independently of the sign of $L^2$, all metrics we consider for $K_p$
are described by Eq.~(\ref{dSmetric}), with $D$ replaced by $p$. Solving
Einstein's equations does not require $K_p$ to be maximally symmetric,
rather it must simply satisfy \reef{BackgroundCurv}. Hence,
for $p>2$, $K_p$ can be taken to be a $p$-dimensional
Schwarzschild-(anti)-de~Sitter solution with:
\begin{equation}
V(r) = 1 -{\mu\over r^{p-3}}- {r^2\over L^2} \,.
\end{equation}
This introduces an additional parameter, the ``mass'' $\mu$, into the
space of solutions.  We will ignore this freedom in the $L^2>0$ case,
where we set $\mu=0$ because empty dS$_p\times S^q$ has the largest
horizon area in that family.  However, in the $L^2<0$ case, we will
find that black holes offer a convenient way of adding unlimited
entropy without affecting the stability of an asymptotically
AdS$_p\times S^q$ solution.

\section{Stability} \label{StableSec}

Naively, many of the solutions discussed in the
previous section appear to violate the $N$-bound. However, to be
confident that we have identified a true violation, we must show
that the solutions are in fact stable. Hence,
we now examine the perturbative stability of these product solutions.
We consider the AdS$_p$ cases
first, generalizing analogous work for $\Lambda = 0$~\cite{DewFre02}.
The computations involve performing a Kaluza-Klein decomposition of
fluctuations on the compact space, and comparing the effective masses
of the resulting modes to the Breitenlohner-Freedman stability bound.
We argue that a black hole can be introduced into the AdS$_p$ without
modifying the stability of the solutions.  Then we consider the
dS$_p$ cases, obtained by continuing the AdS$_p$ results, and identify
instabilities as negative mass-squared modes.  Finally we comment
briefly on the fate of the unstable solutions.

There is a long history of Kaluza-Klein analysis on Freund-Rubin-type
backgrounds, including notably the studies of the fluctuations around
the maximally supersymmetric solutions of supergravity, {\em e.g.}
Refs.~\cite{CasDau84,KimRom85,Van85}; for a review, see
Ref.~\cite{DufNil86}.  The fluctuations on the full space are viewed
as towers of modes on $K_p$, after performing the Kaluza-Klein
reduction on $M_q$.  Our present analysis relies heavily on that
presented in Ref.~\cite{DewFre02}, where the complete fluctuation
spectrum of modes in $K_q \times M_q$ backgrounds
(\ref{BackgroundCurv}), (\ref{BackgroundFlux}) of Einstein-Maxwell
theory (\ref{actpq}) with $\Lambda = 0$ were obtained; the Lorentzian
factor in that case is necessarily negatively curved and was taken to
be anti-de Sitter space.  The form of the fluctuation equations did
not depend on this choice, but the criterion for stability in
principle may: it is well-known that in AdS$_p$, scalar fluctuations
may have negative mass-squared without generating an instability as
long as the masses do not violate the Breitenlohner-Freedman
bound~\cite{BreFre82}:
\begin{eqnarray}
m^2 \tL^2 \geq - {(p-1)^2 \over 4} \,.
\label{bf}
\end{eqnarray}
The existence of this stability window for naively tachyonic modes may
be thought of as a consequence of the negative contribution to the
energy from the negative mass-squared being overwhelmed by the
positive contribution from the spatial variation.

Defining the fluctuations (valid for $p >2$)
\begin{eqnarray}
 \delta g_{\mu \nu} &=& h_{\mu \nu} = H_{\mu \nu}  - {1 \over {p-2}}
 g_{\mu \nu} h^\alpha_\alpha \,, \nonumber \\  \label{MetricAdS}
\delta g_{\mu \alpha} &=& h_{\mu \alpha} \,, \quad \quad \delta
g_{\alpha \beta} = h_{\alpha \beta} \,, \quad \quad\delta A_{q-1} =
a_{q-1} \,, \quad\delta F_q \equiv f_q = da_{q-1} \,, \\
 H_{\mu \nu} &=& H_{(\mu \nu)} + {1 \over p} g_{\mu \nu} H^\rho_\rho \,,
 \quad \quad h_{\alpha \beta} = h_{(\alpha \beta)} + {1 \over q}
 g_{\alpha \beta} h^\gamma_\gamma \,, \nonumber
\end{eqnarray}
with $g^{\mu \nu} H_{(\mu \nu)} \equiv0\equiv g^{\alpha \beta}
h_{(\alpha \beta)}$, it was found for $\Lambda = 0$ \cite{DewFre02} that
all fluctuation equations but one (the one for the modes $h_{(\alpha
\beta)}$) depend on the choice of Einstein spaces solely through the
parameters $p$ and $q$.  Moreover, these modes never violate the
Breitenlohner-Freedman bound.  Among these stable modes are the
diagonal fluctuations of the metric and the fluctuations of the
$q$-form along $M_q$, which form a coupled system that must be
diagonalized.  It was found that the minimum mass $m^2(\lambda)$ of
the Kaluza-Klein modes on the lower branch of this coupled system
exactly saturates the Breitenlohner-Freedman bound (\ref{bf}) without
violating it, as a function of (minus) the eigenvalue of the Laplacian
on the compact space $\lambda$.

In contrast, the fluctuation equation for $h_{(\alpha \beta)}$ depends
on the details of the choice of $M_q$.  It was shown that for $M_q =
S^q$ this field produces only positive-mass modes, and hence a generic
AdS$_p \times S^q$ Freund-Rubin spacetime is perturbatively stable.
In contrast, for the choice $M_q = S^n \times S^{q-n}$ with $q < 9$ an
instability is generated in anti-de Sitter space, and other choices of
$M_q$ may lead to instabilities as well.

Our analysis for nonzero $\Lambda$ uses the same expansion in
spherical harmonics and gauge choices as Ref.~\cite{DewFre02} and
proceeds along the same lines.  Furthermore, the dS$_p$ case can be
obtained simply from the AdS$_p$ case by continuing $\tL^2 \rightarrow
- L^2$.  It is not difficult to show that most fluctuation equations
are unchanged from when $\Lambda = 0$; this includes the $h_{(\alpha
\beta)}$ mode, so once more only positive-mass modes will appear in
this channel as long as the compact space is $S^q$.  Since we are
looking for stable solutions, we will generally take $S^q$ in what
follows.  Then the only potential instability lies in the modes that
are affected by $\Lambda \neq 0$, namely the coupled system mentioned
previously,
\begin{eqnarray}
h^\alpha_\alpha(x,y) &=& \sum_I \pi^I(x) Y^I(y) \,, \quad \quad
H^\mu_\mu(x,y) = \sum_I H^I(x) Y^I(y) \,, \label{GaugeFluct1} \\
a_{\beta_1 \ldots \beta_{q-1}}(x,y) &=& \sum_I b^I(x) \;
 \epsilon^{\alpha}_{\;\;
\beta_1 \ldots \beta_{q-1}} \nabla_\alpha Y^I(y) \,,
\nonumber
\end{eqnarray}
which constitutes a fluctuation of the fields active in the
background.  One still finds (for most cases; see below) that $H^I$
may be algebraically eliminated in favor of $\pi^I$,
\begin{eqnarray}
H^I = - {2 \, (D-2) \over q \, (p-2)}  \pi^I \,,
\label{Heqn}
\end{eqnarray}
leaving the coupled system of equations
\begin{eqnarray}
\label{Coupled}
\tL^2 \square_x \pmatrix{ c b^I \cr \pi^I} = \pmatrix{ \bar\lambda^I &
{2 \over q} (\tilde\alpha + (p-1)(D-2)) \cr {2 q (p-1) \over (D-2)}
\bar\lambda^I & \bar\lambda^I + {2(p-2) \over (D-2)} \tilde \alpha +
2(p-1)^2} \pmatrix{ c b^I \cr \pi^I } \,.
\end{eqnarray}
Here, $\tilde\alpha \equiv 2 \Lambda \tL^2$ is related to the
dimensionless flux ${\cal F}$ by
\begin{eqnarray}
\tilde\alpha = { (D-2)(p-1) \over (q-1) {\cal F} -1} \,,
\end{eqnarray}
and $\bar\lambda^I$ is the rescaled eigenvalue of the compact
Laplacian:
\begin{eqnarray}
-\tL^2 \square_y Y^I \equiv \bar\lambda^I Y^I = \left( {\tL^2 \over
R^2} \right) \lambda^I Y^I \,.
\end{eqnarray}
Diagonalization of the mass matrix (\ref{Coupled}) gives the spectrum
of masses $m^2(\lambda)$ for various values of $\tilde\alpha$,
equivalent to fluxes ${\cal F} > {\cal F}_m = 1/(q-1)$, and
continuation $\tilde{L}^2
\rightarrow - L^2$ gives the spectrum for the de Sitter values $0 \leq
{ \cal F} < {\cal F}_m$.  We examine the stability of these
fluctuations in turn.

\subsection{Stability of AdS$_p \times S^q$}

The mass spectrum for negatively curved $K_p$ is
\begin{eqnarray}
m^2(\bar\lambda)\tL^2 &=& \bar\lambda + \tA \pm \left[\tA^2 + 4
\bar\lambda \tB \right]^{1/2} \,, \label{Mass}
\end{eqnarray}
with the constants
\begin{eqnarray}
\tA = {p-2 \over D-2} \tilde\alpha + (p-1)^2 \,, \quad \quad
\tB = {p-1 \over D-2} \tilde\alpha + (p-1)^2 \,.
\end{eqnarray}
When reducing on $S^q$, $\lambda$ takes values $k(k+q-1)$ with $k \in
{\mathbb Z}_+$.  It turns out that the modes at $k=0, \, 1$ are special
cases for which there is only one physical perturbation, while the
other is a spurious gauge mode related to the existence of conformal
Killing vectors on the sphere.  Separate analysis of these special
cases\footnote{If the compact space is not $S^q$, there is only one
spurious mode ($Y^I$ constant), and the minus branch is defined for $k
\geq 1$.}  leads to the conclusion that the ``minus'' branch of
(\ref{Mass}) is truncated to $k \geq 2$, whereas the ``plus'' branch
is valid for all $k \geq 0$.

One may check that for $\tilde\alpha \rightarrow 0$ our result
(\ref{Mass}) coincides with that of Ref.~\cite{DewFre02}.  Namely, the
minimum occurs on the negative branch for $\bar\lambda_{\rm min} = (3/4)
(p-1)^2$, and at this minimum
\begin{eqnarray}
m^2(\bar\lambda_{\rm min})\tL^2 = - \tf14 (p-1)^2 \,,
\quad \quad (\Lambda = 0)
\end{eqnarray}
which precisely matches the Breitenlohner-Freedman bound for
fluctuations in AdS$_p$.

Since the mass curve (\ref{Mass}) with $\Lambda=0$ saturates the
Breitenlohner-Freedman bound, one might suspect that turning on
$\Lambda$ will push the curve over the edge.  For a fixed AdS$_p$ mass
scale $1/\tL^2$, a positive value of $\Lambda$ causes the compact
space to curve more strongly, and so one might think that the
Kaluza-Klein scale $1/R^2$ set by $M_q$ will become more extreme
relative to $1/\tL^2$.  

Indeed this is what we find.  The mass curve
$m^2(\bar\lambda)$ has a minimum at
\begin{eqnarray}
\bar\lambda_{\rm min} = \tB - {\tA^2 \over 4\tB} \,,
\end{eqnarray}
where it takes the value
\begin{eqnarray}
\label{MassMin}
m^2(\bar\lambda_{\rm min})\tL^2 =
- {(2\tB-\tA)^2 \over 4\tB} = - {\tilde\alpha \over D-2} - {1 \over 4(p-1)} {
\left((p-1)^2 + {p-2 \over D-2} \tilde\alpha \right)^2 \over
\left( (p-1) + {1 \over D-2} \tilde\alpha \right) } \,.
\end{eqnarray}
For infinitesimal $\tilde\alpha$ (very large flux ${\cal F}$), this
becomes
\begin{eqnarray}
m^2(\bar\lambda_{\rm min})\tL^2 \approx - \tf14 (p-1)^2 -
{p+1 \over 4(D-2)} \, \tilde\alpha \,.
\end{eqnarray}
Hence the minimum of the mass curve (\ref{Mass}) dips immediately
below the Breitenlohner-Freedman stability bound as a positive
cosmological constant is turned on.  One may also show that
$m^2(\bar\lambda_{\rm min})\tL^2 $ is monotonically decreasing (so the
minimum mass does not rise above the bound for any larger $\Lambda$),
and that it approaches $ - \infty$ as $\Lambda \rightarrow \infty$.

This analysis suggests that all AdS$_p \times S^q$ solutions with
$\Lambda > 0$ are unstable to small fluctuations.  However, the story
is more subtle.  Recall that only discrete values of $\bar\lambda$
actually arise as eigenvalues of the Laplacian on $M_q$, indexed by
$k$.  As discussed in Ref.~\cite{DewFre02}, for $q$ odd there is
always an integer $k = (q-1)/2$ such that $k(k+q-1)$ is $\lambda_{\rm
min}$, but for $q$ even there is not; the lowest-mass states sit on
either side of the minimum of the curve.  In the former case,
infinitesimal $\Lambda$ will push the lowest state below the
instability bound; this is true even though for $\Lambda > 0$ the
physical state $k=(q-1)/2$ no longer sits precisely at the minimum of
the curve.  In the latter case, however, there is some range with the
cosmological constant sufficiently small in which the lowest-mass
state has not yet moved below the bound.  Hence, for $q$ even and
${\cal F}$ sufficiently large, there is a window of stability.

Furthermore, for $q=2$ and $q=3$, the mode nearest to the minimum
would be at $k=0$ or $k=1$; however, these modes are absent from the
lower branch.  Consequently, one must check whether the physical modes
with $k \geq 2$ become tachyonic; it turns out that they do not, and
as a result there are no instabilities whatsoever in these cases.

Hence a more intricate pattern of instabilities emerges.  For $q=2$ or
$q=3$, the spacetimes are stable regardless of $p$, and for all values
of ${\cal F} \geq {\cal F}_m$.  For $q \geq 4$, a perturbative tachyon
always exists for $q$ odd, but for even $q$ there is a stable
window for ${\cal F}$ sufficiently large.

Let us consider the generalization of the stability analysis of empty
anti-de Sitter space to the case of AdS-Schwarzschild
solutions.\footnote{The black hole itself is taken to have
horizon radius much larger than the $AdS_p$ length $\tilde{L}$, and
hence is classically and semiclassically stable.}  As the Kaluza-Klein
reduction only involves the compact space $S^q$, it can be carried out
equally well in the presence of a black hole, and one obtains
fluctuations with the same masses as in the case of ordinary AdS$_p$
above.  Therefore, the spacetime would only be jeopardized if the
criterion for stability---that is, the Breitenlohner-Freedman bound
(\ref{bf})---was changed by the introduction of the black hole.

The black hole spacetime asymptotically approaches AdS, and the
boundary conditions on fluctuations required by energy conservation
will consequently be unchanged.  Since imposing these conditions
require the mass-squared to exceed the usual
Brei\-ten\-loh\-ner-Freed\-man bound, we see that the presence of the
black hole cannot stabilize previously unstable modes.

Asymptotically AdS spacetimes have been shown to have non-negative
energy, with zero energy coinciding with pure AdS, for supersymmetric
theories with tachyonic scalars satisfying the Breitenlohner-Freedman
bound \cite{GibHul83,Hul83}.  This was generalized to
non-supersymmetric theories for the case of a single scalar in
Ref.~\cite{Bou84}, and it is natural to think that the proof will hold
for the case of multiple scalars.  These results do not, however, rule
out the possibility that the fluctuations of the scalars (which
violate the dominant energy condition~\cite{GibHul83}) could
destabilize the black hole.  The naive extension of Breitenlohner and
Freedman's derivation to the black hole spacetime, with a regularity
condition at the horizon, fails to prove positivity of the energy.
Gubser \cite{Gub00} has conjectured a criterion generalizing the
Breitenlohner-Freedman bound in arbitrary asymptotically AdS
spacetimes.

Here we simply note that a destabilization would be problematic from
the point of view of the $AdS$/CFT correspondence, in which a large
AdS black hole is described by a finite-temperature field theory, with
no known characteristic instability.  If the Breitenlohner-Freedman
bound was raised even infinitesimally with the introduction of a black
hole, the black hole solutions of AdS$_5 \times S^5$ in Type IIB
supergravity and AdS$_4 \times S^7$ in eleven-dimensional
supergravity, and any other solution with $q$ odd, would decay via the
mode that saturates the Breitenlohner-Freedman bound in the case of
empty AdS.  Consequently we do not expect the introduction of a black
hole to produce new instabilities.

\subsection{Stability of dS$_p\times S^q$} \label{stds}

By continuing $\tilde{L}^2 \rightarrow - L^2$, we find the mass
spectrum for the coupled system (\ref{GaugeFluct1}) in the de Sitter
case,
\begin{equation}
\label{Mass2}
m_\pm^2(\bar\lambda)L^2 = \bar\lambda + A \pm\sqrt{A^2+4B\bar\lambda} \,,
\end{equation}
where
\begin{eqnarray}
A&=&{p-2 \over D-2} \alpha - (p-1)^2 = {(p-1)^2 (q-1) \over 1 -(q-1){\cal
F}} \, ( {\cal F} - {\cal F}_s) \,, \label{denote}\\ B&=&{p-1
\over D-2} \alpha - (p-1)^2 = {(p-1)^2 (q-1) \over 1-(q-1){\cal F}}
\, {\cal F} \,, \nonumber \\
\label{Fs}
\cF_s &\equiv& { 1 \over (p-1)(q-1)} \,.
\end{eqnarray}
and $\alpha \equiv 2 \Lambda L^2$, $\bar\lambda = (L^2/R^2) \lambda$.
As the dS$_p$ solutions exist over the range of values $0 \leq {\cal
F} < {\cal F}_m = 1/(q-1)$, $B$ is always positive-definite, while $A$
flips sign at ${\cal F} = {\cal F}_s$.

We are interested in identifying potentially unstable modes in these
spacetimes. There is no Breitenlohner-Freedman bound for dS$_p$
spacetimes, but rather we must simply require that no tachyonic modes
arise in the Kaluza-Klein reduction, as in flat space.  To see this,
consider the local energy density measured by an observer.  Of course,
the effective cosmological constant contributes a fixed energy
density.  The modes associated with tachyonic fluctuations are found
to be exponentially growing or decaying as one approaches the future
or past boundary---see, for example, Refs.~\cite{RacLim66,LimNie67}.  A
short calculation shows that the contribution of the growing modes to
the local energy density increases without bound in these regions of
the spacetime, and hence quickly overwhelms that of the cosmological
constant.  Consequently they cannot consistently be treated as
perturbations of the dS background.  On the other hand, the same
analysis shows there is no such instability for modes with positive
mass-squared, even in the range $0<m^2<(p-1)^2/4L^2$, as the vacuum
energy density dominates the perturbation at all times.  For an
alternate discussion of dS stability, see Ref.~\cite{AbbDes82}.

Just as in the AdS$_p$ case, the Kaluza-Klein reduction on $S^q$
produces only one mode at each $k=0$ and $k=1$, and only for $k \geq
2$ are both branches of (\ref{Mass2}) sampled.  Interestingly, because
$A$ can change sign, whether the $k=0$ mode is on the plus or minus
branch changes as ${\cal F}$ is varied.  Performing the analysis for
the $k=0$ mode, we find the mass
\begin{eqnarray}
m^2_0 L^2 = 2 A \,.
\end{eqnarray}
When ${\cal F} > {\cal F}_s$, this is on the upper branch, while for
${\cal F} < {\cal F}_s$ it moves to the lower branch; the two branches
meet at zero mass for ${\cal F} = {\cal F}_s$ (but there is still only
one massless state at $k=0$).  We notice immediately that for flux
smaller than ${\cal F}_s$, there is always a tachyonic mode constant
on the $S^q$.  In particular, the case with no flux (${\cal F}= 0$) is
always unstable.\footnote{Our result (and, for $p=2$,
Ref.~\cite{BouHaw97b}) refutes the conjecture made in
Ref.~\cite{HubRan02} that the cosmological horizon in dS$_p\times S^q$
spacetimes without flux is classically stable.}

One can show that $k=1$ remains on the upper branch, and that the
upper branch is a monotonically increasing function on the range
$\bar\lambda \geq 0$ taking only positive values.  Therefore, any
instabilities in the region ${\cal F} > {\cal F}_s$ must be
produced by the $k \geq 2$ modes on the minus branch.  We find that
$m_-^2$ has a minimum at
\begin{equation}
\bar\lambda_{\rm min}=B-{A^2\over4B} \,,
\label{lmin}
\end{equation}
at which point
\begin{equation}
m_-^2L^2 =- {(2B-A)^2\over 4B}\,.
\label{mmin}
\end{equation}
Naively it seems that this branch always produces tachyonic modes and
hence all of the corresponding solutions should be unstable.  However,
this need not be so, since $\bar\lambda_{\rm min}$ need not correspond
to any physical excitation.  One must check that there is some integer
$k \geq 2$ that actually produces a tachyon.

One may show that on the range of positive $\bar\lambda$, $m^2_- L^2$
is negative for the region
\begin{eqnarray}
\label{Tachyonrange}
0 < \bar\lambda < \bar\lambda_0 \equiv
{2(p-1)^2 (q-1) \over 1 -{\cal
F} (q-1)} \, ( {\cal F} + {\cal F}_s)\,.
\end{eqnarray}
Since $\bar\lambda \geq 0$, one can see that if a mode with $k \geq 3$
is tachyonic, the $k=2$ mode will be a unstable as well, so the
question of whether there are instabilities in the higher modes
reduces the question of whether $\bar\lambda(k=2)$ lies in the range
(\ref{Tachyonrange}).
One finds that this inequality is only satisfied if $q \geq 4$ and
\begin{eqnarray}
\label{highertachyonbound}
{\cal F} > {\cal F}_t \equiv {2 \over q(q-3)(p-1)} \,.
\end{eqnarray}
The condition for a stable dS$_p \times S^q$ spacetime, incorporating both the $k=0$ tachyon and unstable modes from higher $k$, is thus
\begin{eqnarray}
\label{tachyonbound}
{\cal F}_s \leq {\cal F} \leq {\cal F}_t \,.
\end{eqnarray}
Since no instabilities arise from higher-$k$ modes for $q=2$ or $q=3$,
spacetimes with a compact $S^2$ or $S^3$ are unstable for fluxes
${\cal F} < {\cal F}_s$ only.  The stable region continues through the
Minkowski point at ${\cal F} = {\cal F}_m$ and into the $AdS$ region,
which is also free of instabilities, as remarked in the last
subsection.\footnote{There have been suggestions that the stability
criterion for fluctuations in de Sitter space is more strict than we
assume, in particular that there may be an ``inverse BF bound'' $m^2
L^2 \geq (p-1)^2/4$ for stability \cite{Deb01}.  We note that even assuming
this less favorable criterion, we find stable solutions.}

For $q=4$, (\ref{tachyonbound}) implies a window of stability,
$1/[3(p-1)] \leq {\cal F} \leq 1/[2(p-1)]$.  The instability at $
1/[2(p-1)]< {\cal F}$ continues through the Minkowski region into the
$AdS$ region.

For $q \geq 5$, we find ${\cal F}_t < {\cal F}_s$.  Consequently the
two regions of instability overlap, and instabilities occur for all
values of ${\cal F}$.  Hence spacetimes with $q \geq 5$ can only be
stable in the $AdS$ region, and then only for $q$ even and very large
flux.

One may additionally ask what can be said about the fate of the
unstable solutions.  It would be interesting to apply the analysis of
Ref.~\cite{HorMae01} to this question.  However, this is complicated
by the absence of an asymptotically flat region.  In cosmological
spacetimes, the existence of an apparent horizon does not guarantee
that an event horizon will be present at the initial time where
perturbations are introduced.  Hence it is not clear that the
formation of a dS$_D$ event horizon would require the type of
topological transition that the Horowitz-Maeda argument excludes.
Thus the ultimate fate of the unstable solutions remains an open
question.

\section{Entropy}
\label{EntropySec}

In this section we show that some of the solutions we found have
entropy greater than $N(\Lambda)$.  Hence, they violate the $N$-bound.

Our solutions contain no entropy in the form of ordinary matter
systems---all potential contributions to entropy come from the
Bekenstein-Hawking entropy of event horizons.  In order to interpret
the horizon areas as entropy, furthermore, we must consider only the
stable solutions.  Event horizons are determined by the global
structure of a spacetime, not by its shape at an instant of time.  If
we were to attribute entropy (and hence, temperature) to one of the
unstable product solutions, we would run into a contradiction: the
thermal fluctuations would destabilize the spacetime, and the far
future (including any event horizon) would differ from the assumed
unstable solution.

Let us first consider the stable product solutions with ${\cal F} >
{\cal F}_m$.  At fixed ${\cal F}$, there exist AdS$_p\times
S^q$ solutions with a black hole of arbitrarily large horizon area
($\mu \to \infty$), except if $p=2$.  Hence, there will be a solution
with entropy greater than $N(\Lambda)$.  Thus, the $N$-bound is
violated for all stable Schwarzschild-AdS$_p\times S^q$ solutions
which also satisfy $p>2$.

Next, consider the stable product solutions with ${\cal F} < {\cal
F}_m$.  At fixed ${\cal F}$, the entropy $S(\Lambda, {\cal
F})$ of the corresponding dS$_p\times S^q$ solution is finite.  It is
given by a quarter of the dS$_p$ horizon, times the volume of the
$S^q$:
\begin{equation}
\label{dShor}
S(\Lambda, {\cal F}) = {1 \over 4} \Omega_{p-2} \, \Omega_q \left( {
D-2 \over 2 \Lambda} \right)^{D-2 \over 2} \left( {p-1 \over 1 - (q-1)
{\cal F}}\right)^{p-2 \over 2} \left({q-1 \over 1 + (p-1) {\cal F}}
\right)^{q \over 2} \,.
\end{equation}
This entropy cannot be increased by introducing black holes, because
the cosmological horizon area overcompensates by shrinking.

If the ratio between (\ref{dShor}) and the entropy of the
$d$-dimensional de~Sitter horizon (\ref{general}),
\begin{equation}
{S({\cal F})\over S_0}= {\Omega_{p-2}\, \Omega_q
\over \Omega_{D-2}} \left({p-1\over1- (q-1){\cal
F}}\right)^{p-2\over2} \left({q-1\over1+ (p-1){\cal
F}}\right)^{q\over2}\,,  \label{ratio}
\end{equation}
is greater than unity, the $N$-bound (\ref{Nbound}) will be violated.
The ratio is less than unity for $\cF = 0$, and as $\cF$ increases,
the ratio further decreases until $\cF = \cF_s$.  These
values do not correspond to stable solutions in any case.  Above
${\cal F}_s$ the ratio begins to increase.  It crosses unity at some
value ${\cal F}_{\rm crit}$ and actually diverges at $\cF = \cF_m$.
Hence for a range of values, the ratio is actually larger than one.
For $q=2$ and $q=3$, this entire range corresponds to stable
spacetimes, while for $q=4$, stable spacetimes occur for ${\cal F}_s
\leq {\cal F} \leq {\cal F}_t$.  Consequently there also exist stable
dS$_p \times S^q$ spacetimes that violate the $N$-bound.

Finally, we should remark on the case $p=2$.  The AdS$_2\times S^q$
solutions contain no entropy as such, and we have not found a way of
introducing entropy greater than $N$.  Indeed, we do not expect this
to be possible.  In the vacuum solution, one finds that the area of
the $S^q$ is less than $4N$ independently of the flux.  Introducing
excitations will destroy the asymptotic behavior of the spacetime at
least on one of the two disconnected components of the boundary ${\cal
I}$ \cite{MalMic98}.  Let us consider a solution in which ${\cal I}$
is foliated by a single copy of $S^q$, with area equal to the $S^q$
area of the vacuum solution.  Applying the covariant entropy bound to
this sphere, one concludes that the entropy on the corresponding
light-sheet is less than $N$.  Such light-sheets foliate both the
causal past and the causal future of ${\cal I}$.  Hence, entropy that
would violate the $N$-bound would have to be causally disconnected
from infinity.

The dS$_2\times S^q$ spacetimes are known as charged Nariai solutions.
They correspond (for each value of ${\cal F}$) to the largest charged
black hole in de~Sitter space.  The entropy of such solutions is less
than that of de~Sitter space~\cite{ManRos95}.  The solutions are
classically unstable and develop an asymptotically dS$_{p+2}$ region
only upon perturbation.  Their instabilities have been noted in
Refs.~\cite{GibHaw77a,GinPer83} and studied in detail in
Refs.~\cite{BouHaw96,BouHaw97b,Bou98,Bou99,BouNie99}.

\section{Conclusions}
\label{discuss}

We have shown that some spacetimes with positive cosmological constant
$\Lambda(N)$ contain observable entropy greater than $N$.  Hence they
cannot be described by a theory with a Hilbert space of finite
dimension $e^N$.  We have found no such spacetimes for $D=4$.  The
significance of this exception is not clear.

It would be important to know whether quantum gravity theories with
finite-dimensional Hilbert space exist.  In order to assess this
prospect, one would like to characterize the spacetimes that such
theories would describe.  We have given a number of conditions that
one would expect a suitable class of dual spacetimes, ${\cal C}(N)$, to
satisfy.  A candidate for ${\cal C}(N)$ was proposed in
Ref.~\cite{Bou00b}: the set of spacetimes with positive cosmological
constant, $\mathbf{all}(\Lambda(N))$.

The following conclusions apply for $D>4$:

The product spacetimes we have analyzed demonstrate that the set
$\mathbf{all}(\Lambda(N))$ is not suitable.  It fails to satisfy the
$N$-bound, which demands that none of its spacetimes exhibit
observable entropy greater than $N$.

It follows that the specification of a positive cosmological constant
alone does not suffice to characterize the spacetimes dual to a finite
Hilbert space theory.  There cannot be a straightforward
correspondence between $\Lambda$, and the number of degrees of
freedom, $N$ (a $\Lambda$-$N$ correspondence).

This does not exclude the possibility that a positive cosmological
constant should be properly regarded as a consequence of a
finite-dimensional Hilbert space.  However it cannot be the sole
consequence.

Although we have demonstrated that $\mathbf{all}(\Lambda(N))$ is not
an adequate choice for ${\cal C}(N)$, we have not shown that some
other suitable ${\cal C}(N)$ cannot exist.  All the problematic
examples we found are of product form, and hence cannot be obtained
from dS$^+$ Cauchy data by smooth deformation.  Consequently it is
possible that there is a ${\cal C}(N)$ that excludes these solutions,
yet contains the dS$^+$ solutions and is still closed under
deformations of Cauchy data, without violating the $N$-bound.  Its
characterization remains unclear.

Experience with non-perturbative definitions of string theory has
taught us that the same low energy Lagrangian (in fact, the same
fundamental theory) 
can have different `superselection' sectors described by different
dual theories.  For example, in the AdS/CFT correspondence, the size
of the gauge group of the CFT dual to a ($\Lambda=0$) Freund-Rubin
compactification depends on the flux. In the present context then, it
could be that a particular dual theory ${\cal T}$ with a specific
number of degrees of freedom $N$ will only capture the spacetime
physics of a certain sector of a given low energy theory. In this
case, one approach to defining the corresponding ${\cal C}$ may be to
include fluxes alongside the cosmological constant\footnote{The
specification of flux will identify an isolated sector only if
flux-changing instantons~\cite{BroTei87,BroTei88} are completely
suppressed.  This is the case for the $\Lambda=0$ product solutions
studied in the AdS/CFT context, but one might expect an onset of
non-perturbative instabilities if $\Lambda>0$.} among the parameters
necessary to be specified in order to ensure that the entropy will not
exceed $N$. This appears to produce rather complicated conditions
whose sufficiency is not obvious.  Moreover, this approach fails to
capitalize on the need to deal only with smooth deformations of dS$^+$
spacetimes.

We have argued that a restriction to dS$^+$ spacetimes, where the
$N$-bound is satisfied, is unnatural because of the Cauchy data
deformation problem.  However, it is conceivable that what appear to
be smooth, infinitesimal deformations from the point of view of the
solution space to the effective Lagrangian may in fact be ruled out in
the full non-perturbative quantum theory.

In $D=4$, the set $\mathbf{all}(\Lambda(N))$ is still a viable
candidate for ${\cal C}(N)$.  In general it remains a challenge either
to find, or to prove the non-existence of, a class of spacetimes that
could represent the classical limit of a quantum gravity theory with a
finite number of degrees of freedom.

\section*{Acknowledgments}

RB and RCM are grateful to the CIAR for the opportunity to attend the
Cosmology and Gravity Program's meeting in February 2001, where this
project was initiated. RCM would also like to thank the Institute for
Theoretical Physics at UCSB for hospitality at various stages of this
research, and the organizers and participants of the PASI School on
Quantum Gravity, held on January 4--14, 2002, at the Centros de
Estudios Cient\'\i ficos in Valdivia, Chile, for the opportunity to
lecture on the material presented here.  We would like to thank Tom
Banks, Dan Freedman, Steve Gubser and Gary Horowitz for interesting
conversations.  This research was supported in part by NSERC of Canada
and Fonds FCAR du Qu\'ebec, and by the U.S. National Science
Foundation under Grant No.~PHY99-07949.

\bibliographystyle{board}
\bibliography{all}

\end{document}